\documentstyle[12pt,aasms]{article}
\def\gtsim{\ {\raise-0.5ex\hbox{$\buildrel>\over\sim$}}\ }
\def\ltsim{\ {\raise-0.5ex\hbox{$\buildrel<\over\sim$}}\ }
\slugcomment{To appear in The Astrophysical Journal Letters}

\begin{document}

\title{Dwarf Galaxies Also Have Stellar Halos}

\author{Dante Minniti $^{1,2}$ \& Albert A. Zijlstra $^{2}$}

\altaffiltext{1}{Lawrence Livermore National Laboratory, Livermore, 
CA 94550\\
E-mail:  dminniti@llnl.gov}

\altaffiltext{2}{European Southern Observatory, D--85748 Garching b.
M\"{u}nchen, Germany\\
E-mail: azijlstr@eso.org}

vskip 0.5cm
{\it ApJ Letters, in press}

\begin{abstract}
We present evidence for the existence of an old stellar halo in the
dwarf irregular galaxy WLM, an isolated member of the Local Group.
The halo consists of population-II stars, with low metallicities and
age $\geq 10^{10}$yr.  
The finding of a halo in a dwarf irregular galaxy argues for a
generic mode of galaxy formation that requires a halo in the presence
of  a disk, regardless of galaxy size. This implies that formation mechanisms 
are similar along the spiral Hubble sequence, and favors the scenarios where 
the formation takes place during the original collapse and accretion 
of the protogalactic gas clouds.
Halo formation also appears not necessarily
to be related to the presence of a bulge or a nucleus, since WLM lacks
both of these components.

\end{abstract}

\section{Introduction}

The Milky Way (MW) has an extended halo consisting of old and
metal-poor stars, the so-called Population-II stars (Baade 1944).  A
modern definition of Population II stars would be age $\geq 10^{10}$
yr, and $[Fe/H] \leq -1.0$.  The importance of this halo in terms of
galaxy formation theories was recognized very early on (Eggen et
al. 1962, ELS): the halo is the oldest component of the MW and its
stars must have formed before the MW had fully collapsed.  There are
two models for halo formation: In the original ELS scenario the halo
stars formed while the primordial gas cloud was still
collapsing. Alternatively, if the MW evolved through merging and the
accretion of smaller satellites (as predicted by the cold dark matter
clustering model), the halo would represent the disrupted stellar
component of the original small galaxies, while the disk formed from
the gaseous component (Searle \& Zinn 1978).

The two other large spiral galaxies of the Local Group, M31 and M33,
also have old stellar halos (Mould \& Kristian 1986).  However, the
systematic properties or even existence of such halos along the
morphological sequence of galaxies of Hubble (Hubble 1926) are not
well known.  If the interpretation of the Hubble types in terms of a
formation sequence is correct (Sandage 1986, Kennicutt 1993), the
absence of halos among the least evolved galaxies would be a strong
argument in favour of an origin in mergers, assuming halo stars are 
made in mergers, whereas the presence of
halos would favour an origin in the initial collapse.

The dwarf galaxies are considered to be the least-evolved galaxy type.
They are thought to form from small perturbations in the early
Universe, at the bottom of the hierarchy, which then cluster
gravitationally on progressively larger scales (White \& Rees 1978).
As Dekel \& Silk (1986) conclude: ``The dwarfs (including the so
called ``irregulars'') are actually the common, typical galaxies,
while the ``normal" galaxies are the special cases.''  There are two
types of dwarf galaxies: gas-poor dwarf spheroidals (dSph), and
gas-rich irregulars (dIrr). Most emphasis has been laid on the role of
dwarf spheroidals in galaxy merging. However, dwarf spheroidals are
found primarily in the cores of dense clusters, indicating they have
already undergone significant evolution through interaction (Bingelli
et al.  1987). Also, it is clear that gas-rich spiral galaxies such as
the MW could have not formed by merging of only dSph galaxies.  In
contrast, the dIrr galaxies are rather unevolved, as indicated by
their abundance in low-density regions.  The significance of dIrr
galaxies as possible elements contributing to build larger galaxies
like our own has to some extent been overlooked: it is likely that
they are in fact the major surviving representatives of the original
building blocks. The presence or absence of Population II halos in
dIrr's could help to unveal the formation of the MW.

The closest example of a dIrr is the Large Magellanic Cloud (LMC).
Its proximity ($D = 50$ Mpc) makes it the best studied galaxy after
the MW. It appears that the LMC does not have an extended stellar
halo: the old stars in the LMC have disk-like kinematics (Olszewski
et al. 1996).  The 16 known Population II globular clusters also do
not have kinematics consistent with an isothermal, or slowly rotating,
pressure supported halo (Freeman et al. 1983, Schommer et al. 1992).
However, it is not clear to what extent the very well studied LMC can
be used as a typical dwarf due to strong past interactions with the MW
and with the Small Magellanic Cloud (Lin, \& Lynden-Bell 1977, Murai
\& Fujimoto 1980).  Is the absence of stellar halos a common feature
of dIrr's (Gallagher \& Hunter 1984), or is it just that the LMC halo
has been perturbed, or even stripped away, by interactions with the
MW?

A few of the numerous dIrr's known in the Local Group of galaxies are
distant enough from other major galaxies to have remained unperturbed
during a Hubble time. The Im IV--V galaxy WLM (DDO~221) is one such
system: lying about a Mpc away from the dominant massive spirals, it
is considered to be a member of the Local Group (van den Bergh 1994).
The closest known galaxy to WLM is the Irr V galaxy IC 1613 (DDO~8),
at a separation of $\Delta D \sim 0.37$ Mpc.  {}From least action
principle considerations (Peebles 1995), it is very likely that WLM
has remained isolated throughout its lifetime.  
Thus, WLM is a good example of
a galaxy at the extreme of galaxy formation.  Note also that
WLM has only one, old globular cluster (Ables \& Ables 1977, Sandage
\& Carlson 1985), in contrast to the Magellanic Clouds which have a
very rich system of luminous, young clusters.

\section{Metallicities, Ages, and Distribution of Stars in WLM}

Deep V and I images of WLM and neighboring fields were obtained under
photometric conditions with the red arm of EMMI at the NTT telescope
in La Silla, which was operated in remote from the ESO headquarters in
Garching (Minniti \& Zijlstra 1996).  Two of these frames are shown in
Figure 1 (Plate 1).

Figure 2 shows the number density of red stars ($20.5 < I < 22.5$,
$V-I > 0.7$) as function of distance from the major axis obtained with
DAOFIND.  The radial position is defined as $r = [x^2 + (y
0.41$) (Ables \& Ables 1977).  The inclination of WLM is 69$^{\circ}$
(Ables \& Ables 1977), and there is no evidence of a warp in the
disk of WLM from the HI maps (Huchtmeier et al. 1981).  Most of the
stars are concentrated in the disk of the galaxy, within $\sim$2
arcmin of the major axis. However, there is a low density tail of red stars
extending as far as the edge of the field shown in Figure 1. Figure 2
proves that there is a faint halo around WLM: while the disk extends
to about 2.5 arcmin from the WLM minor axis, the halo can be traced
out to about 2 kpc (the scale is 0.264 kpc arcmin$^{-1}$).  The
surface-density profile of this halo follows a de Vaucouleurs law,
although the radial extent is not large enough to formally exclude
other fits.  Counting stars allows us to reach fainter levels than
possible with traditional surface-photometry methods (Pritchet, \& van
den Bergh 1994).  We explore the halo of WLM down to very faint
equivalent surface-brightness levels ($\mu_V \approx 30$ mag
arcsec$^{-2}$). We also measure the flattening of the WLM
halo, $b/a = 0.6$, using two other CCD fields located to the S and W of
these shown in Figure 1. Thus, this halo is significantly rounder than the
central population.

Colour--magnitude diagrams for the main body and the outer parts of
WLM were obtained with DAOPHOT~II (Figure 3).  The appearance of these
colour--magnitude diagrams show a clear population difference between
the inner and the outer parts of the galaxy: they can be understood in
terms of a superposition of an extended old and metal-poor population
plus a younger, more concentrated component, adding scatter introduced
by possible differential reddening and by star crowding in the inner
regions. This distribution is indicative of a halo-like component in
WLM, with either a lower metallicity and/or a larger age than
the central disk population.

To prove that this extended stellar population is similar to the MW
halo, an age of $\gtsim 10^{10}$yr would have to be established.
Usually, metallicity and age are degenerate, meaning that they cannot
be discriminated with two-colour photometry. Fortunately, in the case
of WLM there is an important additional piece of information: Skillman
et al. (1989) measured $Z = 0.001$ in the HII regions in the WLM disk.
Since the gas reflects the cumulative process of enrichment along the
lifetime of a galaxy, it is safe to assume that the stars are not more
metal-rich than the gas in the WLM disk.  The high C/M star ratio
(Cook et al. 1986) in the WLM disk confirms that the intermediate-age
stellar population is also very metal poor. 
We will therefore assume that the halo is metal poor as well.

With this assumption, the distance to this galaxy can be
determined. For $[Fe/H] \leq -1.0$, the tip of the red-giant branch
(RGB) has constant $M_I = -4.0 \pm 0.1$, independent of metallicity
(Lee 1993). From Figure 4 we find $I_{RGBT} = 20.80 \pm 0.05$, which
is in excellent agreement with the value reported by Lee et al. (1993)
$I_{RGBT} = 20.85 \pm 0.05$.  For $A_I = 0.10$, the distance modulus
to WLM is $m-M = 24.80 \pm 0.05$, which translates into $D = 0.91$
Mpc.

The metallicity of the WLM halo can be directly measured from
the color and shape of the RGB, if the distance is known. Using recent
calibrations (Lee 1993, Da Costa, \& Armandroff 1990), we obtain
[Fe/H]$ = -1.4\pm 0.2$ from the V$-$I colour of the RGB at I$ =
$I$_{RGBT}+0.5$.  This is similar to the abundance of the WLM disk
(Skillman et al. 1989), so that the halo and disk of this galaxy show
comparable metallicities. The difference of appearance of disk and
halo (Figure 3) must therefore be due to a different age structure.

The age of the WLM halo stars can in principle be derived from the
absolute magnitudes of the brightest stars on the asymptotic giant
branch (AGB).  The observed magnitude of the AGB tip is $I_{AGBT} =
19.90 \pm 0.10$ (Figure 4). Using recent isochrones (Bertelli et
al. 1994) we derive an age of about $10^{10}$ yr for the WLM halo
(Figure 3). The precise age depends on colour corrections used for
obtaining the bolometric magnitudes, and on the assumptions about
overshooting and mass loss, which are presently not well known
(Bertelli et al. 1994). Independent of these assumptions, there is a
jump in the luminosity function at the location of the RGB tip that is
more than a factor of 4, indicative of a very old population (Renzini
1992).  Clearly the WLM halo is old, consistent with he presence of a
globular cluster in WLM. However, this age estimate is uncertain and
needs to be confirmed, for example, by deeper HST photometry, reaching
the horizontal branch.

\section{Discussion}

The age and metallicity which we derive for the old stars in WLM are
consistent with earlier results (Ferraro et al. 1989, Lee 1993). The
disk contains stars from both components, whereas the halo lacks the
young stars.  This population difference is in strong contrast with
the LMC, where old and young stars have very similar spatial
distributions.  This suggests that in WLM the two populations have
different formation scenarios.
 
Da Costa (1994) pointed out that all galaxies of the Local Group,
regardless of type and luminosity, have Population II stars. This
conclusion was based on the presence of globular clusters, RR Lyrae
variables, or blue horizontal branches and metal-poor giants, with the
only possible exception to the rule being Leo II (Da Costa 1994).  In
the case of the dIrr galaxy NGC~3109, a field away from the disk also
has old and metal-poor stars (Lee 1993).  The presence of halos in
dwarf galaxies may then be the rule rather than the exception.

Direct evidence for old halos is now available for four galaxies
including WLM, and possibly for NGC3109.  The WLM halo shows that
Population II halos are not a unique characteristic of large spiral
galaxies. The formation of the WLM halo is not only linked to the
evolution process (e.g. by mergers) of large spiral galaxies. This
finding favours the scenario where the formation of a halo takes place
during the original collapse of the proto-galactic gas cloud: initial
perturbations of all amplitudes form stars from the primordial gas
shortly after the recombination era, with the rest of the gas
collapsing into the inner regions forming stars at a later epoch. In
addition to the halo formation being independent of spiral-galaxy
type, it also appears not be related to the presence of a bulge or a
nucleus, since WLM lacks both of these components.  This supports the
suggestion that the bulge and the halo of the MW are different
components (Minniti 1996). It also implies that dIrr galaxies are the
natural continuation of the Hubble disk sequence.

The mere presence of stellar halos and Population-II stars formed at z
$\geq 4$ contradicts some scenarios where the formation of dwarf
galaxies is delayed until $z \leq$ 1 due to the photoionization of the
gas by the metagalactic EUV radiation (Babul \& Rees 1992). However,
this mechanism could still have acted to halt star formation in WLM
after the halo formation, producing a large gap before star formation
took place in the disk (Ferraro et al. 1989). The similar metallicity
of gas and halo stars may indicate that star formation has
been intermittent only.

In addition to the now-discovered old stellar halos, dIrr galaxies are
known to have dark matter halos, their total M/L being larger than for
normal spiral galaxies (C\^ot\'e 1995).  In the case of WLM, the
observed amplitude of the HI rotation curve $\sim 50$ km/s (Huchtmeier
et al. 1981) also indicates a high M/L. Whether these dark halos are
related to the stellar halos is not clear from this, but remains a
possibility.

If WLM were accreted by the MW at early times ($z \geq 2$), the
stellar halo of this dIrr galaxy would have been assimilated by the
Galactic halo, while the gaseous component would have sunk to the
central regions of the disk and bulge, contributing to later epochs of
star formation. Thus, the accretion of smaller satellites may still
have contributed to the formation of the MW halo, but we conclude that
in general stellar halos are a product of earlier episodes 
of galaxy formation.

\section{Summary}

We have studied the structure of WLM, in particular examining the
spatial distribution of blue (young) and red (older) stars, and
establishing the presence of an extended halo, down to very faint
levels.  {}From the VI colour--magnitude diagram we infer metallicities
and ages for the stars in this halo, proving that it is made of
Population-II stars.  {}From the deep luminosity function we obtain an
accurate distance for this galaxy.  Using this information, we give an
indication of the age of the WLM halo, which turns out to be typical
for Population-II stars.

Thus, WLM represents a case where no spiral arms, nucleus
or bulge are present, but where a disk formed dissipatively within an
old and metal-poor halo.

\acknowledgements{We are very grateful to S. White, E. Skillman, and
A. Renzini for many suggestions and comments. We thank
M. Kissler-Patig for help with DAOPHOT~II, and S. C\^ot\'e,
E. Olszewski, and M. Mateo for copies of their works, and the
anonymous referee for useful suggestions.  This work was performed in
part under the auspices of the U. S. Department of Energy by Lawrence
Livermore National Laboratory under Contract W-7405-Eng-48.  }

\clearpage

\begin{figure}
\caption{
CCD I frames of WLM taken under photometric conditions
with the NTT with $\leq 1$ arcsec seeing.
The total field covered is 15.4$\times$8.9 arcmin. North is up, and 
East to
the left.  
A few background galaxies can be seen in the frames.
This field covers most of the light of the galaxy included within the
outer contours of the surface photometry of Ables \& Ables (1977).
}
\end{figure}


\begin{figure}
\caption{
Number density of stars as
function of radial distance from the major axis. The different symbols 
correspond to star counts in 3 different CCD frames. Most stars are
concentrated in the disk, within 2$^\prime$ of the major
axis. However, a low-density tail is seen extending to $8^\prime$ from
the major axis, doubling the known extension of WLM.
}
\end{figure}
\begin{figure}
\caption{
Color--magnitude diagrams of the outer and inner parts of WLM 
(left and right panels, respectively.).
Only stars with centroids
matched in V and I frames to better than 1.5 pixels (0.4"), and
satisfying stringent criteria of photometric quality ($\sigma \leq 0.3$,
$\chi \leq 2$, and Sharpness $\geq -1$) are plotted.
The main features
include a red tail of AGB stars (mostly C stars), a blue main sequence,
a red supergiant sequence. Signs for a population of ``blue loopers" are
also seen, running parallel to the main sequence, about 0.4 mag redder.
}
\end{figure}

\begin{figure}
\caption{
Luminosity functions of all the stellar objects in
the outer halo regions of WLM.
Two distinct features can be seen:
a sharp break at $I = 20.8$ due to the termination of the 
halo--like RGB, and another break at $I = 20.0$
due to the termination of the AGB.  The halo fields are located at
$x > 3$ arcmin, where x is distance away from the major axis.
The optical surface brightness
at $d\geq3$ arcmin is $\leq26$ mag arcsec$^{-2}$ (Ables \& Ables 1977).
After background contamination has been removed, the counts at $I =
20.8$ jump by more than a factor of 4, indicating a very old age ($>
10^{10}$ Gyr) for the WLM halo (Renzini 1992).
}
\end{figure}

\end{document}